\documentclass[preprint,showpacs,preprintnumbers,amsmath,amssymb]{revtex4-1}
\usepackage{graphics,epsfig,subfigure}
\usepackage{color}

\begin{document}
\renewcommand{\baselinestretch}{1.3}
\newcommand\beq{\begin{equation}}
\newcommand\eeq{\end{equation}}
\newcommand\beqn{\begin{eqnarray}}
\newcommand\eeqn{\end{eqnarray}}
\newcommand\nn{\nonumber}
\newcommand\fc{\frac}
\newcommand\lt{\left}
\newcommand\rt{\right}
\newcommand\pt{\partial}

\title{Excited Kerr black holes with  scalar hair}
\author{  Yong-Qiang Wang\footnote{yqwang@lzu.edu.cn
},
 Yu-Xiao Liu\footnote{liuyx@lzu.edu.cn} and  Shao-Wen Wei\footnote{weishw@lzu.edu.cn}
}

\affiliation{Research Center of Gravitation $\&$ Institute of Theoretical Physics  $\&$ Key Laboratory for Magnetism and Magnetic of the Ministry of Education, Lanzhou University, Lanzhou 730000, China}

\begin{abstract}

In the context of complex scalar field coupled to Einstein gravity theory,
we present a novel family of solutions of Kerr black holes with excited-state scalar hair  inspired  by the work of Herdeiro and
Radu in [Phys.\ Rev.\ Lett.\  {\bf 112}, 221101 (2014)], which can be regarded as numerical solutions of
 rotating compact objects with excited scalar hair, including boson stars and black holes.
In contrast to Kerr black holes with  ground state scalar hair,   we find that the first-excited   Kerr black holes with scalar hair have  two   types of nodes,  including  radial $n_r=1$  and angular  $n_\theta=1$ nodes. Moreover,  in the case of radial nodes the curves of the mass  versus the
 frequency  form  nontrivial loops,   and in the case of angular nodes the curves can be divided into two kinds:  closed   and open loops.  We also study the dependence of the horizon  area    on  angular momentum   and   Hawking temperature.
\end{abstract}

\maketitle

\section{Introduction}\label{Sec1}
One of the most interesting  discoveries  in the study of black holes (BHs) is
uniqueness theorem \cite{Chrusciel:2012jk},
 which states that the four-dimensional, asymptotically flat black hole solutions of the
Einstein field equations  is Kerr-Newman black hole (BH).  The theorem is only  valid  for  the Einstein-Maxwell theory,
however,
when more general types of matter fields  are coupled to gravity,  a few counterexamples to uniqueness theorem have been found during the last few years. For instance,
the first example of  black hole  with non-Abelian  hair was discovered  in four-dimensional $SU(2)$ Einstein-Yang-Mills
theory \cite{Colored1,Colored2}. Soon after that,   the black holes with Skyrme hair were found in the non-linear sigma-model of Skyrme coupled to gravity \cite{Droz:1991cx} and  with Yang-Mills-Dilaton  hair  in Einstein Yang-Mills-Dilaton theory \cite{H1}.

 Recently, a family of   Kerr black holes with scalar hair  (KBHsSH)  were discovered by Herdeiro and
Radu  in the four-dimensional Einstein gravity coupled to a complex  scalar field \cite{Herdeiro:2014goa}.  The  scalar hair  is a free massive  field, moreover,
 it takes a  time-dependent harmonic   value in which the time variation is sinusoidal
 and     the frequency $\omega$ of the complex field needs to obey  the synchronisation condition
\begin{equation}
\omega =\Omega_H m,
\end{equation}
where  the constant $\Omega_H>0$ is the horizon angular velocity
and $m$ is the azimuthal harmonic index.    The solutions of   Kerr black holes with scalar hair can   reduce to spinning boson
stars (BSs) in the limit of vanishing horizon area.
The stability of   Kerr black holes with scalar hair was  discussed in Refs. \cite{Ganchev:2017uuo,Degollado:2018ypf} and the bound scalar hair can be regarded as the  zero mode of the
superradiant instability \cite{Hod:2012px,Benone:2014ssa}. In Refs.  \cite{Herdeiro:2014jaa,Cunha:2015yba}, the properties of ergosurfaces  and shadows  were  also  investigated.  Furthermore,  the   study of hairy black holes can be extended
to the Proca hair case  \cite{Herdeiro:2016tmi},  Kerr-Newman BH \cite{Delgado:2016jxq},   non-minimal coupling case \cite{Herdeiro:2018wvd}, and  spinning BHs with Skyrme hair \cite{Herdeiro:2018daq}.
Especially, considering the model of gravity coupled to self-interacting scalar field \cite{Herdeiro:2015tia,Herdeiro:2016gxs},     one can obtain the    solution of  Kerr black holes with ultra-light scalar  hair.
The study of relevant astrophysical observational signatures of   Kerr black holes with scalar hair has been investigated in Refs. \cite{Ni:2016rhz,Cao:2016zbh,Shen:2016acv,Zhou:2017glv}.
With the study on long-term numerical evolutions of  the superradiant instability of  Kerr black hole by East and Pretorius \cite{East:2017ovw},
the relationship between formation properties of
Kerr black hole with Proca hair and superradiant instability of  Kerr black hole was discussed in Refs. \cite{Herdeiro:2017phl,East:2018glu}.
Besides, a variety of   analytical  studies on  hairy black holes have been undertaken  in  Refs. \cite{Hod:2013zza,Hod:2014baa,Hod:2017kpt,Hod:2016yxg,Hod:2017rmh,Hod:2016lgi,Brihaye:2018woc}, and there have been a lot of  attention to study    hairy black holes
recently \cite{Brihaye:2016vkv,Herdeiro:2015moa,Brito:2015pxa,Herdeiro:2017oyt,Herdeiro:2017fhv,Herdeiro:2017fhv,Delgado:2018khf}.
See Ref. \cite{Herdeiro:2015gia} for a detailed discussion of numerical method  and
Ref.  \cite{Herdeiro:2015waa}  for a review.

Until now, only Kerr  black holes  with the scalar hair
in the ground state have been considered, that is,   the  scalar hair   can keep sign along the radial $r$ direction.
On the other hand, in the astrophysical applications of boson stars \cite{Schunck:1997FE,Yoshida:1997qf,Schunck:1996he},   the authors of Refs. \cite{Bernal:2009zy,Collodel:2017biu} found some    boson star
solutions with  excited scalar field  which  were beneficial to obtain more realistic rotation curves of
spiral galaxies. So, it will be interesting to see whether  there are  solutions of Kerr  black hole  with excited-state scalar hair in   Einstein-Klein-Gordon  theory.  In the present paper,  we would like to numerically solve the system of field equations and
give   a  family of   Kerr black holes with  first-excited state scalar hair,
which can be divided into two categories  of   radial nodes  $n_r=1$ and angular nodes  $n_\theta=1$.

 The paper is organized as follows. In Sect. \ref{sec2}, we introduce  the model of
  the four-dimensional Einstein gravity coupled to a free, complex  scalar field  and adopt the same axisymmetric metric with Kerr-like coordinates as the  ansatz in Ref. \cite{Herdeiro:2014goa}.  In Sect. \ref{sec3}, the boundary conditions  of excited  Kerr  black holes  with  scalar hair  and rotating boson stars are studied.
  We show the numerical  results of the equations of motion and study the characteristics of radial nodes   and angular nodes  in Sect. \ref{sec4}.  The conclusion and discussion are given in the
  last section.

\section{The model setup}\label{sec2}

Let us begin with the model of  (3+1)-dimensional Einstein gravity coupled to a free, complex massive scalar field,  with
the Lagrangian density
\begin{align}\label{EKG}
\mathcal{L} =\frac{R}{16\pi G}-\nabla_a\psi^*\nabla^a\psi-\mu^2|\psi|^2\,,
\end{align}
where $G$ is the gravitational constant and the term proportional to $\mu^2$  is known as a mass term.  The equations of motion of the scalar  field   are given by
\begin{equation}
\Box\psi=\mu^2\psi\,,
\label{eq:EKG2}
\end{equation}
and Einstein field equations read as
\begin{equation}
R_{ab}=8\pi G(\nabla_{a}\psi^*\nabla_{b}\psi+\nabla_{b}\psi^*\nabla_{a}\psi)+8\pi G g_{ab}\mu^2\psi^*\psi.
\label{eq:EKG1}
\end{equation}
Note that  if the  complex, massive  scalar field $\psi $ vanishes, the solution of Einstein equations (\ref{eq:EKG1}),  which can describe  the stationary axisymmetric
asymptotically flat  black
hole with mass  and angular momentum, is the well-known Kerr black hole.
In terms of Boyer-Lindquist coordinates, the Kerr metric reads
\begin{eqnarray}
 ds^{2} &=& - \left(\frac{\Delta - a^2 \sin^2\theta}{\varrho}\right) dt^2
             - \left(\frac{2 a \sin^2 \theta (r^2+ a^2 -\Delta)}{\varrho}\right) dt d \phi\nonumber\\
   &&+ \left(\frac{(r^2+a^2)^2 - \Delta a^2 \sin^2
        \theta}{\varrho}\right) \sin^2 \theta d \phi^2 + \frac{\varrho}{\Delta}\ dr^2
        + \Sigma\ d\, \theta^2 ,
\end{eqnarray}
with  $\varrho = r^{2} + a^{2}\cos^{2}\theta$  and  the Kerr horizon function  $\Delta = r^{2}-2Mr +a^{2}$.
Here, the constants $a$  and $M$ are the angular momentum per unit mass  and  the mass of the Kerr BH  as measured
from the infinite boundary, respectively. The non-extremal
Kerr black hole has the event horizon $r_{+}$ and the Cauchy
horizon $r_{-}$ at
\begin{equation}
r_{\pm}=M \pm \sqrt{M^2-a^2}.
\end{equation}
 The Hawking
temperature $T_H$ and angular velocity $\Omega_H$ of  the Kerr black hole are given by
\begin{eqnarray}
T_H  &=& {r_+ - r_-\over 4\pi(r_+^2+ a^2)}, \nonumber\\
\Omega_H &=& \frac{a}{r_+^2 + a^2}.
\end{eqnarray}

When   there exists a non-trivial  configuration of the complex scalar field,
Herdeiro and Radu   constructed a class of   Kerr BH solutions with ground state scalar hair  \cite{Herdeiro:2014goa}.
In order to construct   stationary   solutions of  a Kerr BH with excited state scalar hair,
we also adopt the axisymmetric metric with Kerr-like coordinates \cite{Herdeiro:2014goa,Herdeiro:2015gia} within the following ansatz
\begin{eqnarray}
\label{ansatz1}
ds^2=e^{2F_1}\left(\frac{dr^2}{N }+r^2 d\theta^2\right)+e^{2F_2}r^2 \sin^2\theta (d\varphi-W dt)^2-e^{2F_0} N dt^2,
\end{eqnarray}
with $N=1-\frac{r_H}{r}$ and the constant $r_H$ is  related to 
event horizon radius.
In addition,  the ansatz of the complex matter field is given by
\begin{eqnarray}\label{ansatz2}
\psi=\phi_n e^{i(m\varphi-\omega t)}, \;\;\;  n=0,1,\cdots,\;\;\; m=\pm1,\pm2,  \cdots .
\label{scalar_ansatz}
\end{eqnarray}
Here, $\varphi$ is the azimuthal angle  and  the  five functions $F_{i}~(i=0,1,2)$, $W$ and $\phi_n$ depend on the radial distance  $r$  and  polar angle $\theta$.
The constant $\omega$ is the frequency of the complex scalar field and $m$ is the azimuthal harmonic index. Subscript  $n$ is  named as the principal quantum number of the scalar field, and  $n=0$ is regarded as the ground state and $n\geq1$ as the excited states.

By demanding  the
Euclidean space smooth  and continuous at the horizon, 
one can determine the Hawking temperature with the metric (\ref{ansatz1})
\begin{eqnarray}
\label{THAH}
T_H=\frac{1}{4\pi r_H}e^{F_0(r_H, \theta)-F_1(r_H, \theta)}.
\end{eqnarray}
The Bekenstein-Hawking  entropy associated with the  horizon is given by
\begin{equation}\label{entropy}
  S=\frac{A}{4G}=\frac{\pi r_H^2}{2 G}\int^\pi_0 d\theta \sin \theta~e^{F_1(r_H, \theta)+F_2(r_H, \theta)}.
\end{equation}

The field   equations  (\ref{eq:EKG2}) and (\ref{eq:EKG1})  with the ansatzs (\ref{ansatz1}) and (\ref{ansatz2}) are a set of  seven non-linear coupled partial differential equations (PDEs).  One of the seven equations which comes from the scalar field equation (\ref{eq:EKG2}) is a second-order PDE for the function $\phi_{n}(r, \theta)$.  With linear combinations of the components of Einstein field equations, the remaining six equations from Eq. (\ref{eq:EKG1})  yield a new set of PDEs,  including  four equations for finding  solutions and two constraint equations for checking the numerical accuracy.
 It is not convenient to write down these seven equations in this paper, and see Ref. \cite{Herdeiro:2015gia}
 for more details.  With  numerical methods for solving these equations of motion,
we could obtain two classes of solutions:
horizonless boson star (BS) solutions  with $r_H=0$ and hairy black hole solutions with $r_H >0$.  The Solitonic solutions can be seen as   deformations 
of the global  Minkowski spacetime, while the hairy black hole solutions closely correspond to  deformations of the Kerr BH.

It is well known that the ground state scalar hair has no node,  that is,
along the radial $r$ direction, the value of the scalar field has the same sign.
When studying   Kerr black holes with the excited  scalar hair,
we find that there are two types of nodes, including   radial and angular nodes.    Radial nodes are the points where
 the value of  the scalar field can  change sign along the radial $r$ direction, while,  angular nodes  are the points where
 the value of the scalar field can  change sign along the angular $\theta$ direction.

\section{Boundary conditions}\label{sec3}
Before  numerically solving the differential equations instead of seeking the analytical solutions, we should obtain the asymptotic behaviors of the  five functions $F_{i}~(i=0,1,2)$, $W$ and $\phi_n$, which is equivalent to give the boundary conditions we need.
Considering  the  properties of
Kerr black holes with excited state scalar hair, we will still use  the  boundary conditions by following the same steps as  the ground state   given in Refs. \cite{Herdeiro:2014goa, Herdeiro:2015gia}.

Considering an axial symmetry  system, we have polar angle  reflection symmetry $\theta\rightarrow\pi-\theta$ on the equatorial plane, and thus  it is convenient to consider the
 coordinate range $\theta \in [0,\pi/2] $. So,  we require  the functions to satisfy the following boundary conditions  at $\theta=\pi/2$
\begin{equation}
\partial_\theta F_i(r, \pi/2) = \partial_\theta W(r, \pi/2) =\partial_\theta \phi_n(r, \pi/2) = 0,
\end{equation}
and  set  axis boundary conditions at $\theta=0$ where  regularity must be imposed    Dirichlet boundary conditions on
$\phi$ and Neumann boundary conditions on the other functions
\begin{equation}\label{abc}
\partial_\theta F_i(r, 0) =\partial_\theta W(r, 0)=\phi_n(r, 0)=0.
\end{equation}
In addition,
the asymptotic behaviors near the boundary $r\rightarrow\infty$ are
\begin{equation}\label{rbc}
  F_i=W=\phi_n=0.
\end{equation}
And finally, 
by expanding the equations of motion near
$r = r_H$ as a power series in $(r-r_H)$, we have
 \begin{equation}
\partial_r F_i(r_H, \theta)   = 0,
\end{equation}
and 
\begin{eqnarray}
 \partial_r \phi_n(r_H, \theta) = 0, \nonumber\\
  W(r_H, \theta) =\frac{\omega}{m}
\end{eqnarray}
for  black hole solutions with  $r_H>0$, and 
\begin{eqnarray}
 \phi_n(0, \theta) = 0,\nonumber\\
  \partial_r W(0, \theta)  = 0
\end{eqnarray}
for boson star solutions with $r_H=0$.
Note that    the values of  $W(0, \theta)  $ and  $F_i(0, \theta)  $ are the constants   independent of   the  polar angle $ \theta$.

Properties of the black hole  can be obtained from the asymptotic behavior of the solutions.
Near the boundary  $r\rightarrow\infty$, the metric functions  $g_{tt}$ and $g_{\varphi t}$     have the following forms
\begin{eqnarray}
\label{asym}
g_{tt} \rightarrow-1+\frac{2GM}{r}+\cdots, \nonumber\\
g_{\varphi t}\rightarrow-\frac{2GJ}{r}\sin^2\theta+ \cdots,
\end{eqnarray}
where the parameters $M$ and  $J$  are the mass and  angular momentum   of the hairy Kerr black hole, respectively.
\section{Numerical results}

In this section, we will solve  the above coupled equations (\ref{eq:EKG2}) and  (\ref{eq:EKG1}) with the ansatzs (\ref{ansatz1}) and (\ref{ansatz2}) numerically.
It is convenient to introduce a new coordinate $ x \equiv \frac{\sqrt{r^2-r_H^2}}{1+\sqrt{r^2-r_H^2}}, $
which can  compactify the radial coordinate $r$ and implies that $r = r_H$ at  $x = 0$ and $r = \infty$ at  $ x = 1$. Thus the inner and outer boundaries of the shell
are fixed at $x = 0$ and  $x = 1$, respectively.
After solving the equations numerically,  we can study the dependence on the frequency $\omega$, the scalar field mass $\mu$ and  event horizon $r_H$, respectively.   Due to scalar invariance,
we can work at  a  fixed  scalar field mass. Moreover, for simplicity,  we choose  $G= 1$.

Next,  we will discuss  the principal quantum number $n=1$  of the Kerr BH with scalar hair,  which is  the case of the first-excited state, and show two classes of  radial   and angular node solutions, respectively. The numerical data files for the sample of reference
solutions for excited Kerr BHs with scalar hair  are included in an ancillary file.

\subsection{ Radial nodes  $n_r=1$}
Along the angular $\theta$ direction, the value of the scalar field $\phi_1$ has the same sign for the case of radial node $n_r=1$. However,  along the radial $r$ direction, the  scalar field $\phi_1$  changes sign once  at  some point which is called  radial node. So,  we name this case as  the first bound excited state with radial node, signed with $n_r=1$. In the following,  we will show the results of excited boson star and
Kerr BH with scalar hair,  respectively.

\subsubsection{Boson star}
\begin{figure}[h!]
\begin{center}
\includegraphics[height=.24\textheight]{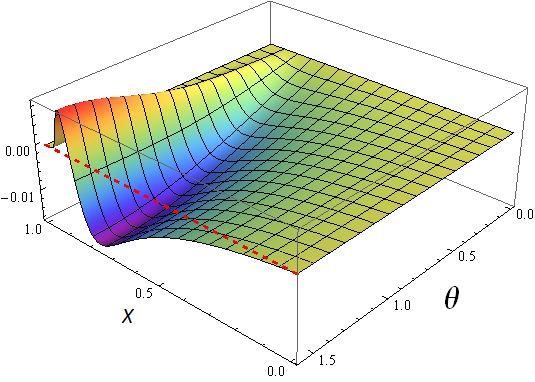}
\includegraphics[height=.24\textheight]{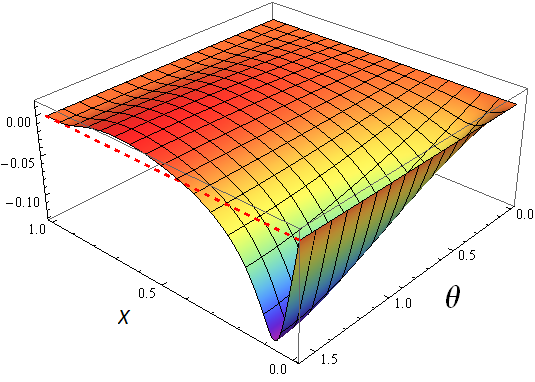}
\end{center}
\caption{Two numerical solutions of  the scalar field $\phi_1$ as a
function of $x$ and $\theta$  for azimuthal harmonic index $m=1$ and frequency  $\omega=0.9$. The red dashed lines   represent  zero value.  }
\label{f-11}
\end{figure}
\begin{figure}[h!]
\begin{center}
\includegraphics[height=.25\textheight]{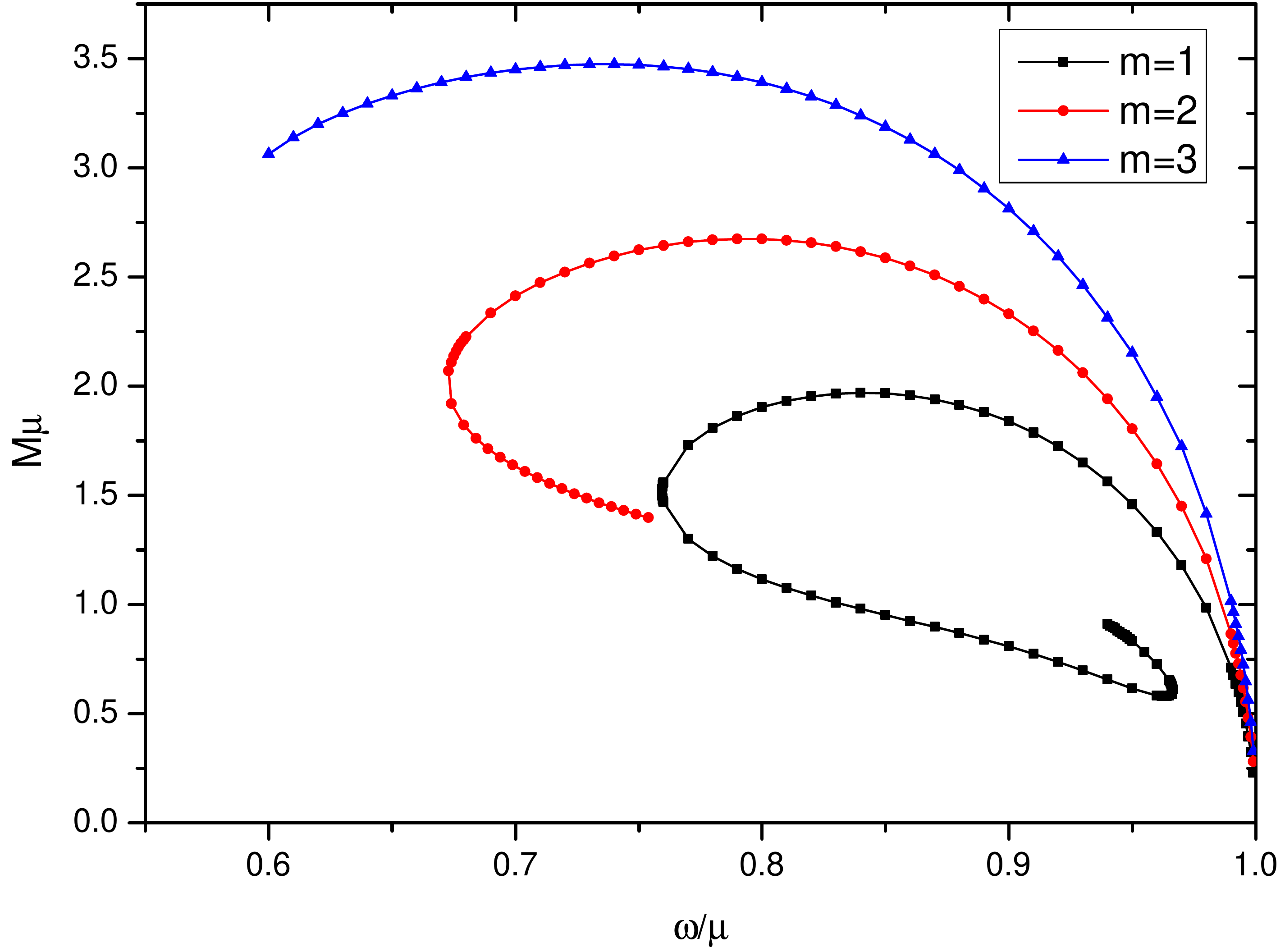}
\includegraphics[height=.25\textheight]{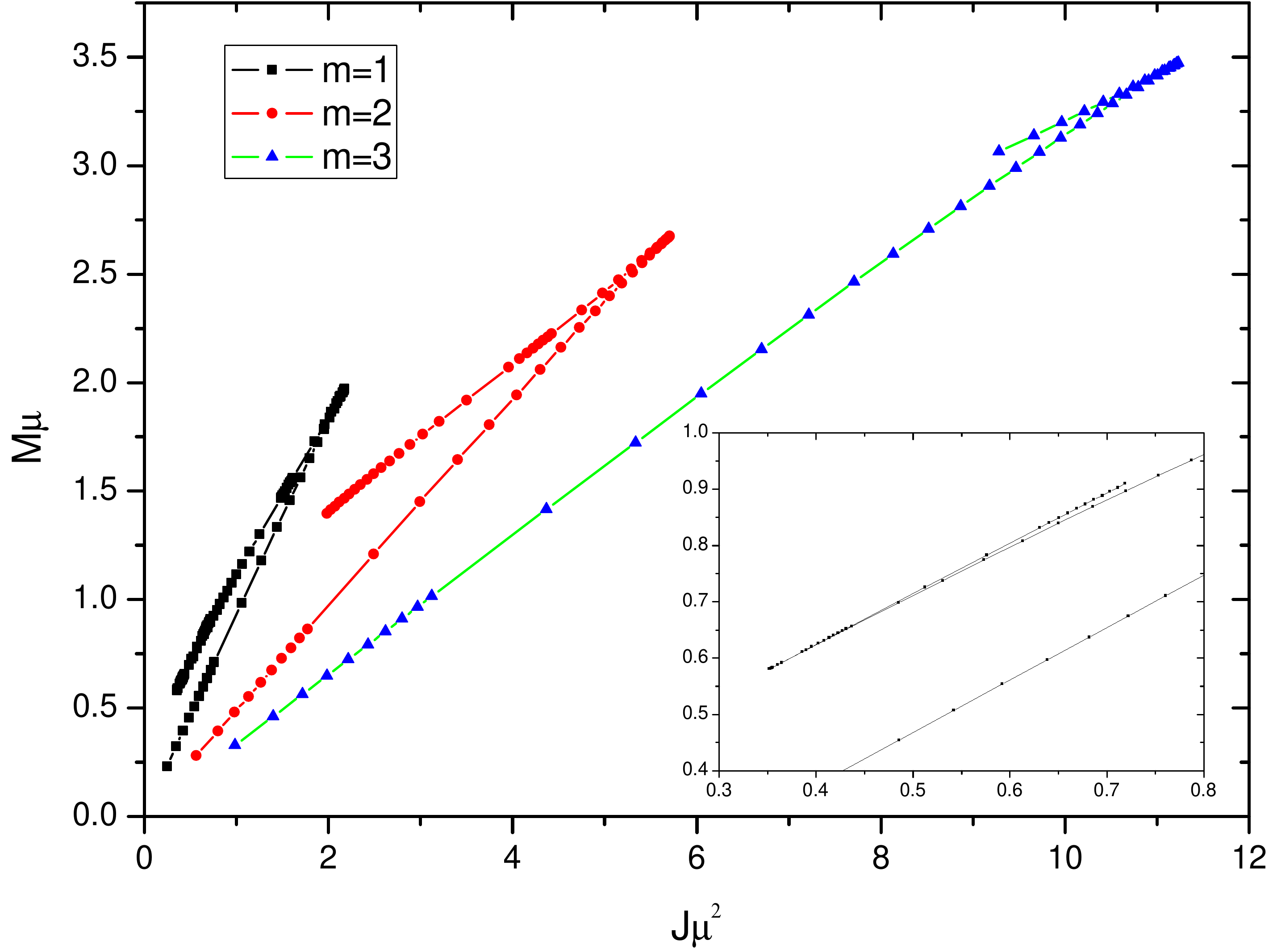}
\end{center}
\caption{The physical property  of the rotating   $n_r=1$  excited state of boson star with $m=1,~2,~3$.  Left graph: the BS mass  $M$ as a function of the  frequency  $\omega$.
Right graph: the BS mass  $M$  as a function of the angular momentum $J$.}
\label{f-12}
\end{figure}

Firstly, as an example, we show in  Fig. \ref{f-11}  two typical results of our numerical program for
the scalar field $\phi_1$ as a function of $x$ and $\theta$ with  $m=1$ for the same parameter $\omega=0.9$.
Along the equatorial plane   at $\theta = \pi/2$, we can observe that  the  scalar field $\phi_1$  changes sign once from the center of the boson star to the boundary  in a  node where there is zero value. In both graphs  the function $\phi_1$ has even parity and could keep the same sign  for the angular variable $0\leq\theta\leq \pi$.
In the left panel of Fig. \ref{f-11},  the distribution of the scalar field is  near the  boundary of spacetime, while the right panel shows
the distribution   is near the  center of the boson star.   Though both  plots have the same parameter,
the first graph belongs to  a   branch of solutions with higher  black hole mass , and the  second graph belongs to  a second set of unstable solutions with lower black hole mass. These behaviors are
further shown in Fig. \ref{f-12}.

To study the properties of the rotating excited state boson star,  in the left panel of  Fig. \ref{f-12} we exhibit   the mass $M$
versus the  frequency  $\omega$  for  three sets of boson stars with the azimuthal harmonic index $m=1~,2,~3$, represented by the  black, red  and  blue lines, respectively,  and  plot  the mass $M$  as a function of the angular momentum $J$   for the corresponding  values of  $m$ in the right panel of  Fig. \ref{f-12}. 
 Note that in this paper all physical quantities are expressed
in units set by   $\mu$.
We can see  from the left plot that there exist   excited state  boson stars   for $\omega < \mu$,  which
means the  excited state solutions are still  bound states   and  similar to the ground state  boson star in Ref. [15].
The spiral curve with $m=1$ starts from the vacuum and  revolves into a central region of the graph. Moreover, when the curve spirals
 into the center, the numerical error begins to increase and  a finer mesh is required to calculate. For simplify, we only show the part of  the curve for $m=2, 3$, which  have similar behaviour as that of $m=1$.
Comparing with the ground BS $\omega-M$    curve in Ref.  \cite{Herdeiro:2014goa}, we can see that the mass of the excited BS is more heavier than that of the ground state,   and  the minimum value  of $\omega$  of the excited state is  larger than that of  the ground state.

On the right panel of Fig. 2, we show
how  the BS mass  $M$ varies as a function of the angular momentum $J$  with  different  azimuthal harmonic index  $m = 1$ (black lines),  2 (red lines), and   3 (green lines).  These three  zigzag  patterns are  exactly similar to that  for the ground state in Ref. \cite{Herdeiro:2014goa},  where the inset shows the detail of the zigzag curves with $m=1$.

\subsubsection{Black hole}
\begin{figure}[h!]
\begin{center}
\includegraphics[height=.25\textheight,width=.34\textheight, angle =0]{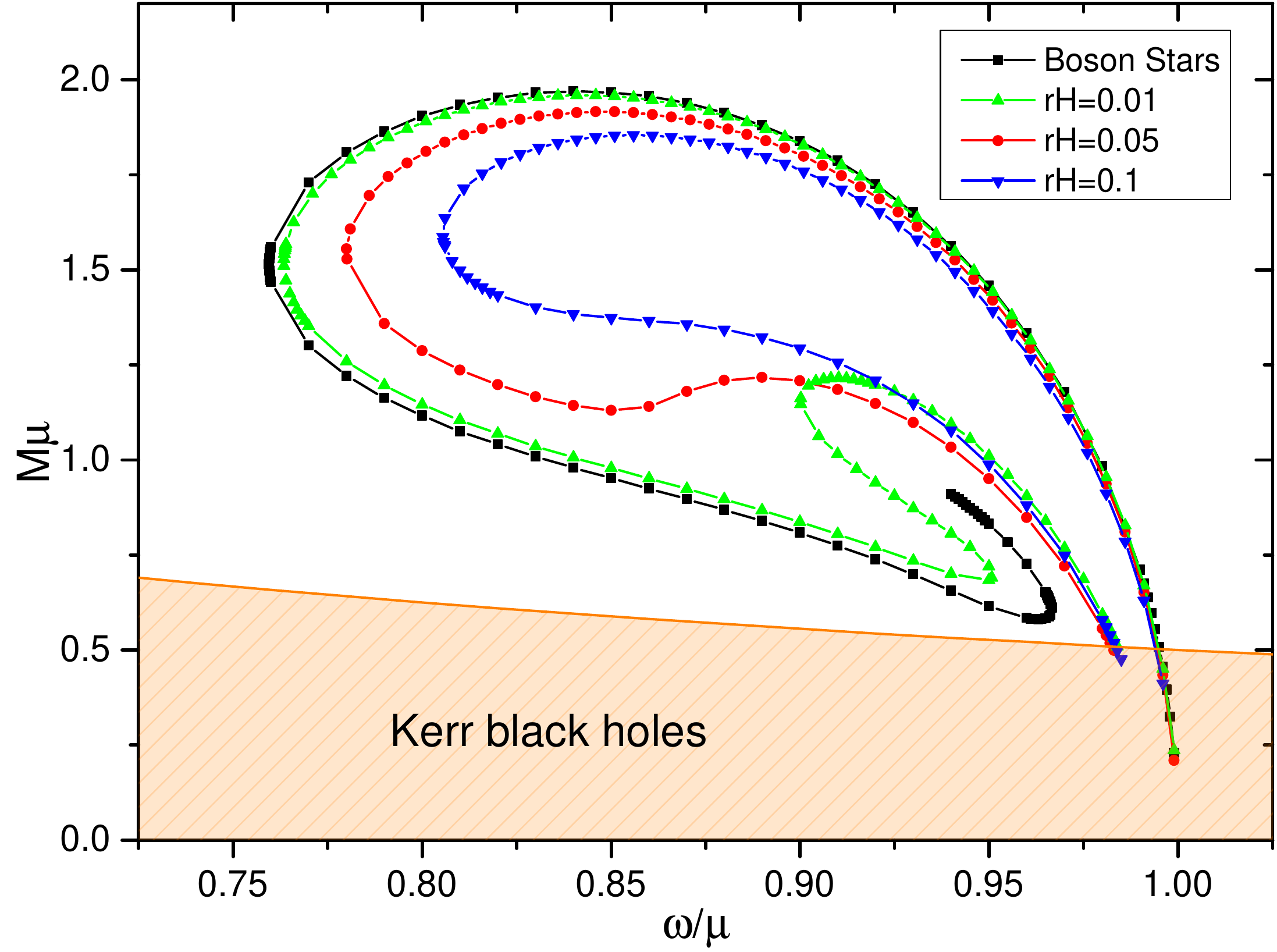}
\includegraphics[height=.26\textheight,width=.34\textheight, angle =0]{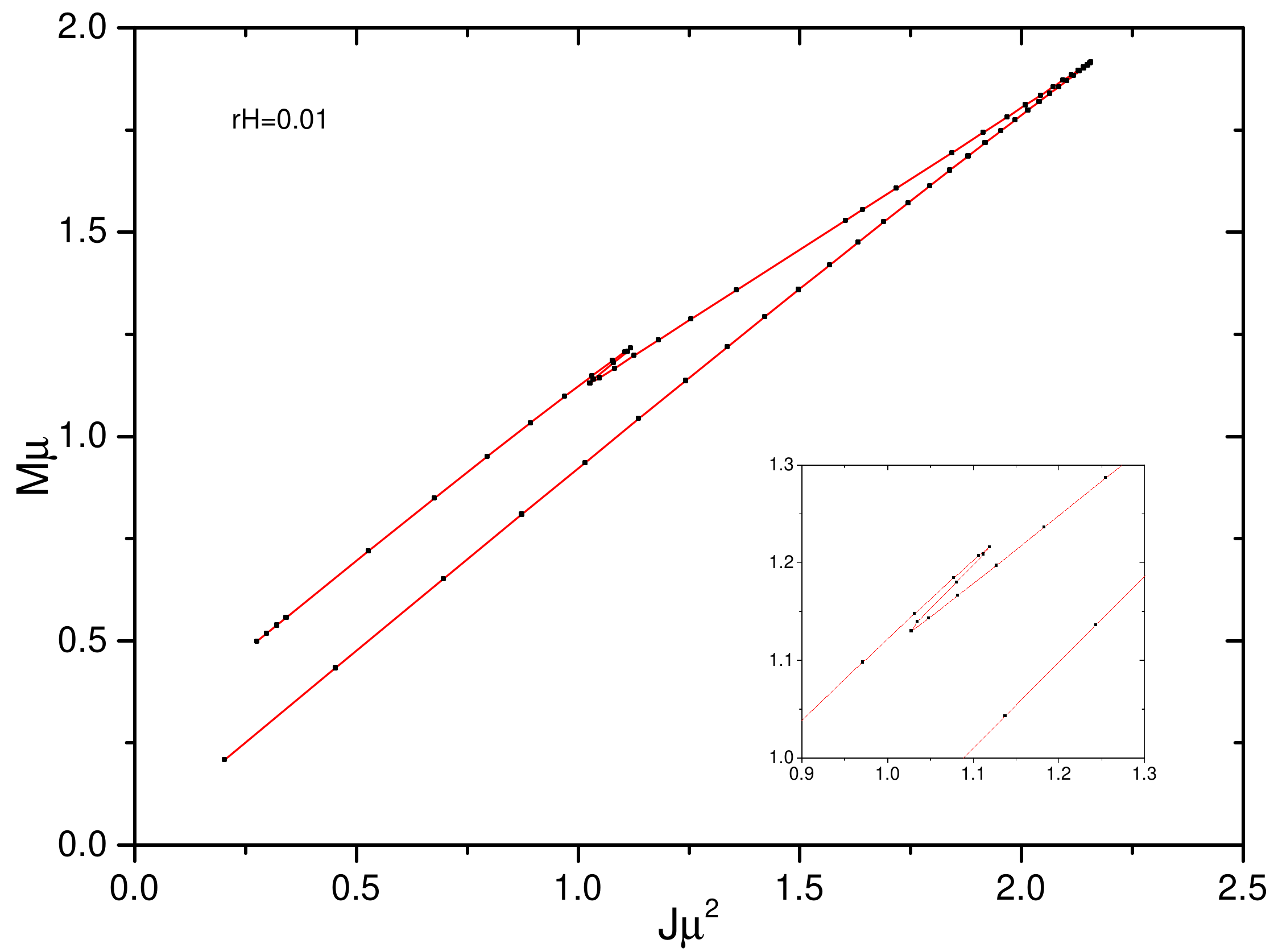}
\end{center}
\caption{The physical property  of  the  $n_r=1$  excited state of Kerr black holes with $m=1$ scalar hair.  Left graph: the BH mass  $M$ as a function of the  frequency  $\omega$  with  the event horizon $r_H=0.01$ (green), $ 0.05$ (red)  and  $0.1$ (blue),   respectively.
Right graph: the BH mass  $M$  as a function of the angular momentum $J$  with  the event horizon $r_H=0.01$.}
\label{bh-2}
\end{figure}

In order to obtain the excited Kerr black hole with scalar hair,  we require the  conditon
$r_H>0$.  In the left panel of Fig. \ref{bh-2}, we  plot  the  mass $M$  of the  excited KBHsSH as a function
of  the  frequency  $\omega$  with $m=1$.  The black line represents the excited BS  curve which has been discussed in Fig. \ref {f-12}.
We plot three curves  with  the event horizon $r_H=0.01$ (the green lines ), $r_H=0.05$ (the  red  lines),  and $r_H=0.1$ (the  blue  lines),   respectively,  
and the domain of existence of Kerr BHs is indicated by the orange shaded area.

With the decrease of the  frequency  $\omega$, the mass of  the excited KBHsSH  increases firstly and then it reaches a maximum point.
Further decreasing $\omega$,  the mass begins to decrease until a minimum value  of $\omega_{min}$  below which   no excited KBHsSHs are found.  As the    frequency  continues to increase to a maximum value,  we obtain a second set of solutions with lower mass.
It is interesting that  the curve  does not form a spiral.  Instead,
they form a closed loop.
As the value of the radius $r_H$ decreases to zero,   the curve of the excited  Kerr BH with scalar hair star
is   very close to the boson star curve.
Here,  we  only exhibit in   Fig.  \ref{bh-2}  the solution of the $m=1$
mode,  and the other values of $m$  have similar behaviour.
In the right panel of Fig. \ref{bh-2}, we  show the mass $M$  as a function of the angular momentum $J$  for $r_H=0.01$.
We find  that the  multi-zigzag  pattern is more complex than that of the ground case,   and more details  are  shown in the  inset.

\begin{figure}[h!]
\begin{center}
\includegraphics[height=.26\textheight,width=.33\textheight, angle =0]{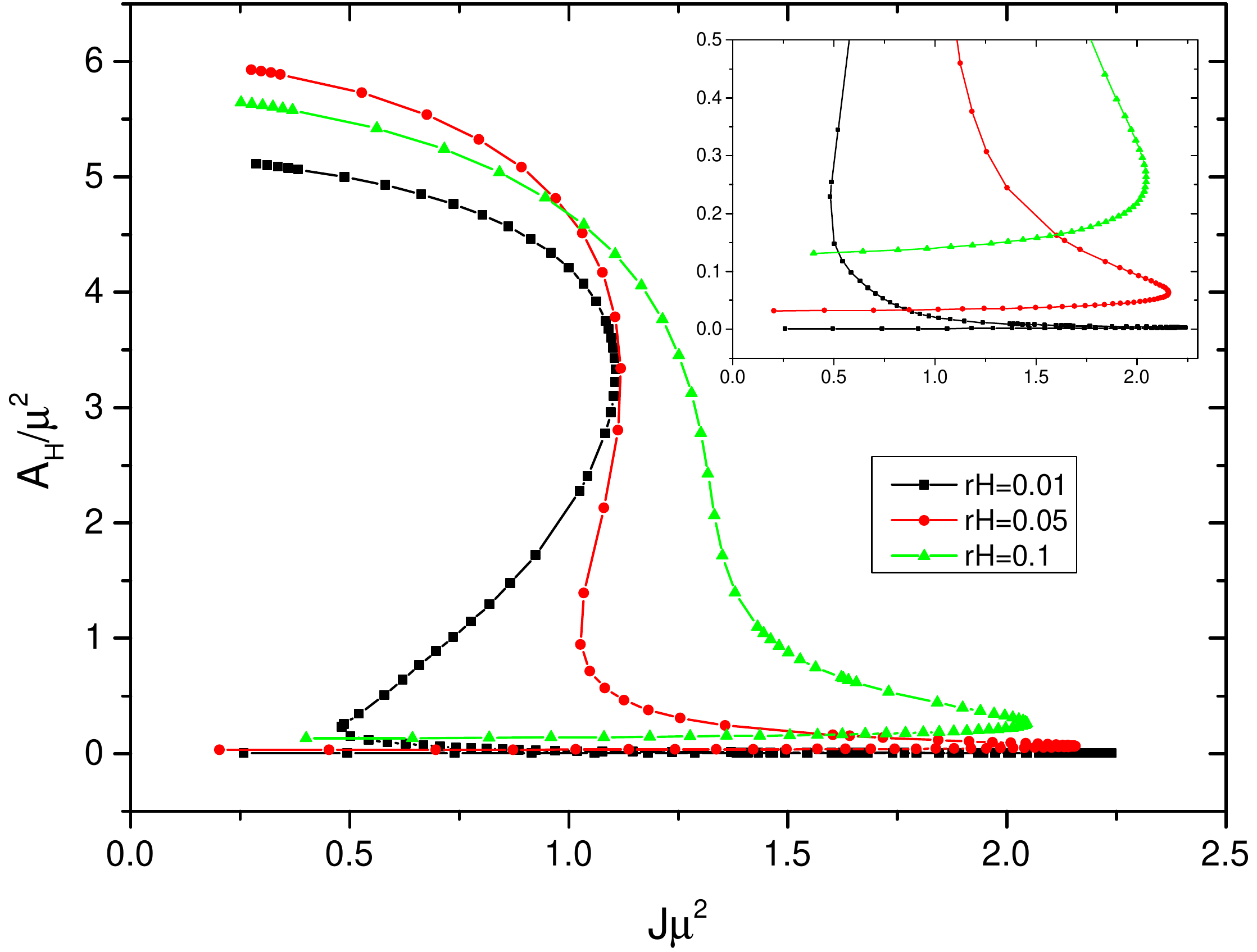}
\includegraphics[height=.26\textheight,width=.33\textheight, angle =0]{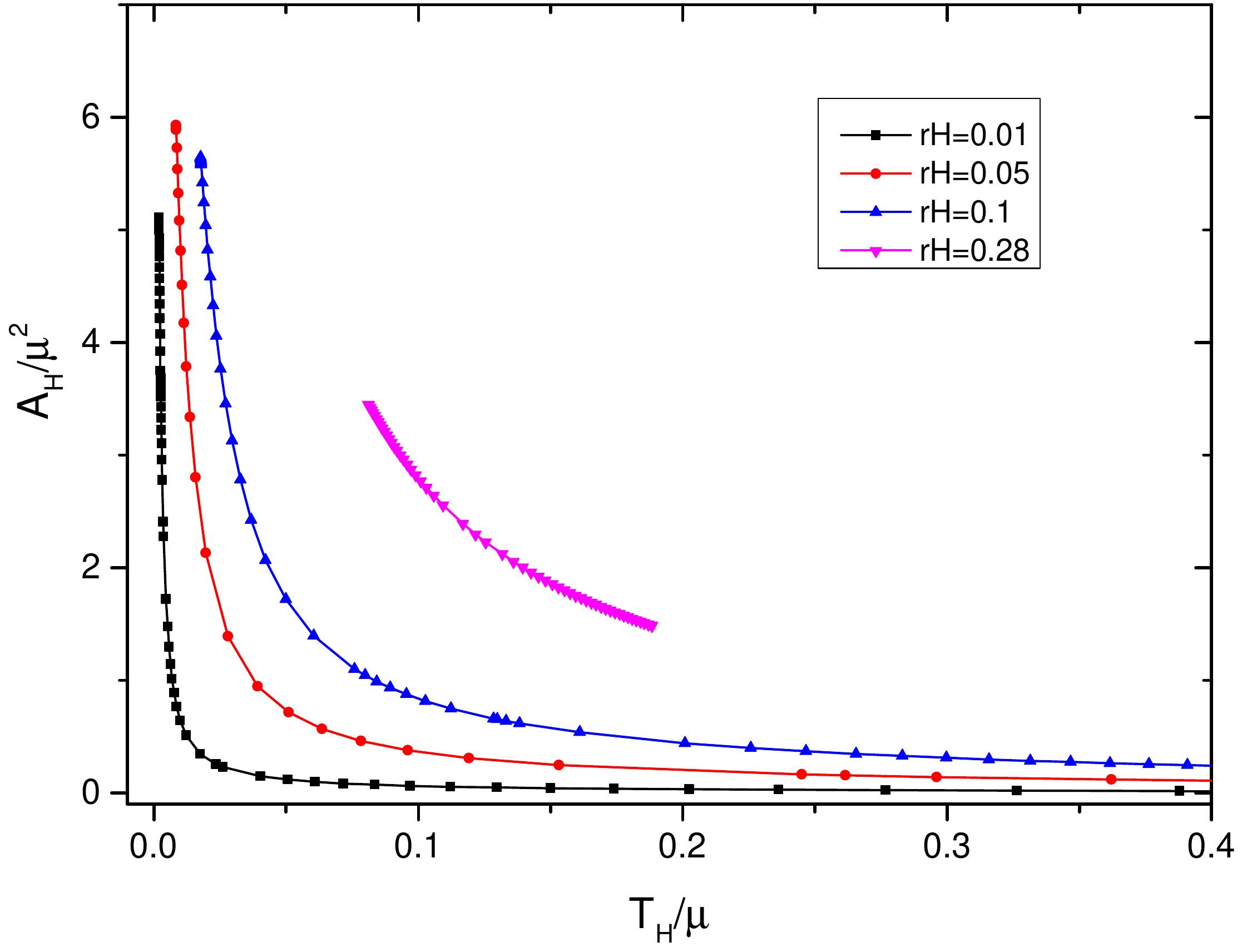}
\end{center}
\caption{The surface area  $A_H$  of the   $n_r=1$  excited state of Kerr black holes with $m=1$ scalar hair.  Left graph:  the surface area   as a function of the angular momentum $J$  with  the event horizon $r_H=0.01$ (black),  $0.05$ (red)  and  $0.1$ (green),   respectively.
Right graph: the surface area   as a function of the Hawking temperature $T_H$,    the parameters of curves from top to bottom are  $r_H$ = 0.28, $r_H$ = 0.1, $r_H$ = 0.05 and $r_H$ = 0.01.}
\label{bh-3}
\end{figure}

As typical examples for our numerical results,   we show  in Fig. \ref{bh-3}   the area $A_H$ of the event horizon   as functions of  the  angular momentum $J$  and temperature $T_H$, respectively,
by taking $r_H=0.01$ (black), 0.05 (red),  and 0.1 (green). In the left panel
 we can  see  that  the  black hole solution with  a fixed radius  has the maximal angular momentum  $ J_{max} $ and there are no hairy  black hole solutions  for  $ J>  J_{max} $. There exist two  curve branches   of  which    the area $A_H$ of the event horizon  has different behaviors with  $J$.  With the decreasing of $J$,
the area of  one  branch decreases  and the other   increases. Moreover,  for the  large area branch,   when the size of $r_H$   decreases,  one
obtains a new third set of  branch in which  the area of the hairy BH is
an increasing function of  $J$. 
 The  large area branch of the hairy BH  is   no longer a monotone   function of  $J$. Indeed, this behavior is in good agreement with the numerical results in Fig. \ref{bh-2}.
More details are given in the inset plotted in the left panel of
Fig. \ref{bh-3}.
In the right panel of Fig. \ref{bh-3},    we can see that for a fixed temperature $T_H$,  the area $A_H$   increases  with $r_H$.
Moreover,
when the event horizon  of the  hairy black hole  decreases,
the temperature $T_H$ can have a wider range.  As an example in Fig.\ref{bh-3},   we show
 the range of the temperature $T_H$, which is  between 0.077 and 0.18  for the pink  line  with  radius $r_H=0.28$.

\subsection{Angular nodes $n_\theta=1$}
In the last subsection,   we gave a family of   boson star and hairy black hole solutions with the  $n_r=1$ excited state,  which means the value of  the scalar field $\phi_1$ has the same sign along the angular $\theta$ direction.  Next,  we will show   other numerical results in which  the value of  the scalar field $\phi_1$ can    change the sign along the angular $\theta$ direction.  In addition,
 along the radial $r$ direction, the  scalar field $\phi_1$ can keep sign or only change sign once  at  some  radial nodes,  which still belongs to  the first bound excited state.  So, we  denote  this family of   solutions with $n_\theta=1$.

\subsubsection{Boson star}
\begin{figure}[h!]
\begin{center}
\includegraphics[height=.23\textheight]{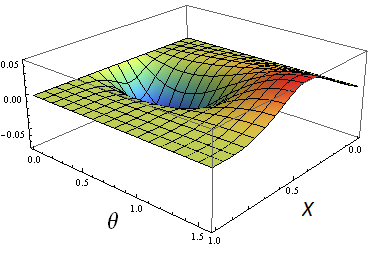}
\includegraphics[height=.23\textheight]{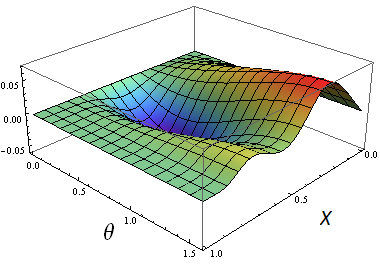}
\includegraphics[height=.23\textheight]{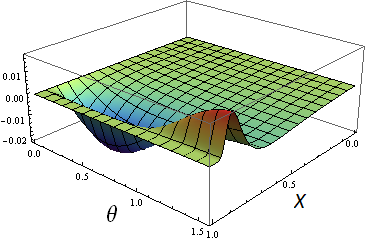}
\includegraphics[height=.23\textheight]{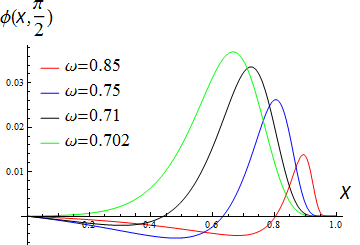}
\end{center}
\caption{  The spatial profiles of   three typical numerical results  for the scalar field $\phi_1$  of rotating BSs with $m=1 $ and $\omega=0.77$.   In the bottom right panel,  we show the distribution of the scalar field as a function of  the $x$ coordinate  in  the equatorial plane  at $\theta = \pi/2$.}
\label{f-b1}
\end{figure}

To describe the  properties of  the first bound excited state, we show the spatial profile of   three typical numerical results  for
the scalar field $\phi_1(x, \theta)$ with  $m=1$  in  Fig. \ref{f-b1}. These three plots have  the same  angular momentum  $\omega=0.77$.
 The sign of the  scalar field $\phi_1$ could change along the angular $\theta$ direction.  However,
along the radial $x$ direction,    the  scalar field $\phi_1$   changes sign in the  bottom left   panel  or keeps sign in the top  panel.
These three plots correspond to three branches of boson star solutions, respectively.
As an example,   the distribution of the scalar field as a function of  the $x$ coordinate for different values of the angular momentum  in  the equatorial plane  at $\theta = \pi/2$   is shown in the bottom right  panel of Fig. \ref{f-b1}.

\begin{figure}[h!]
\begin{center}
\includegraphics[height=.26\textheight,width=.34\textheight, angle =0]{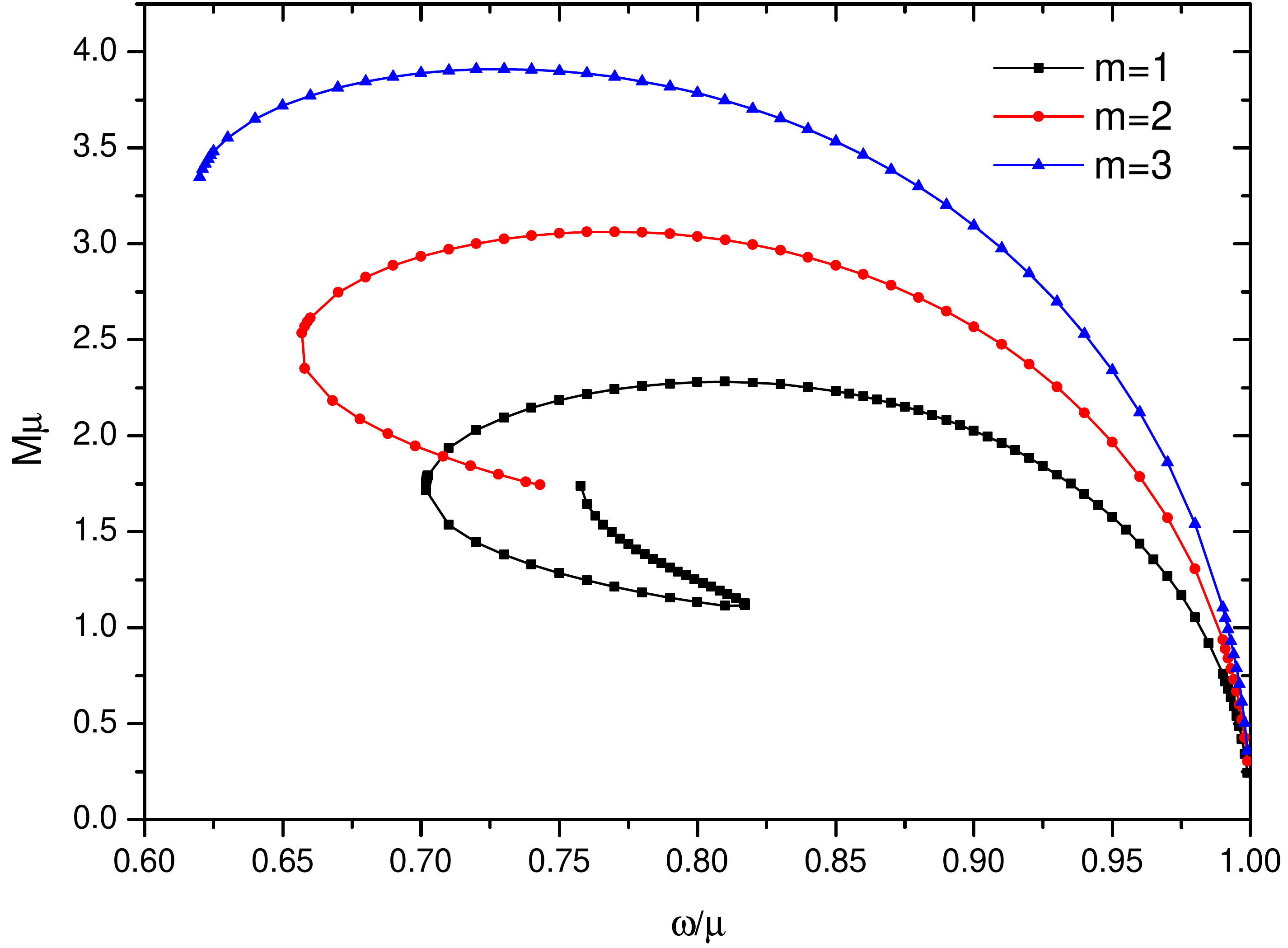} \ \ \
\includegraphics[height=.26\textheight,width=.34\textheight, angle =0]{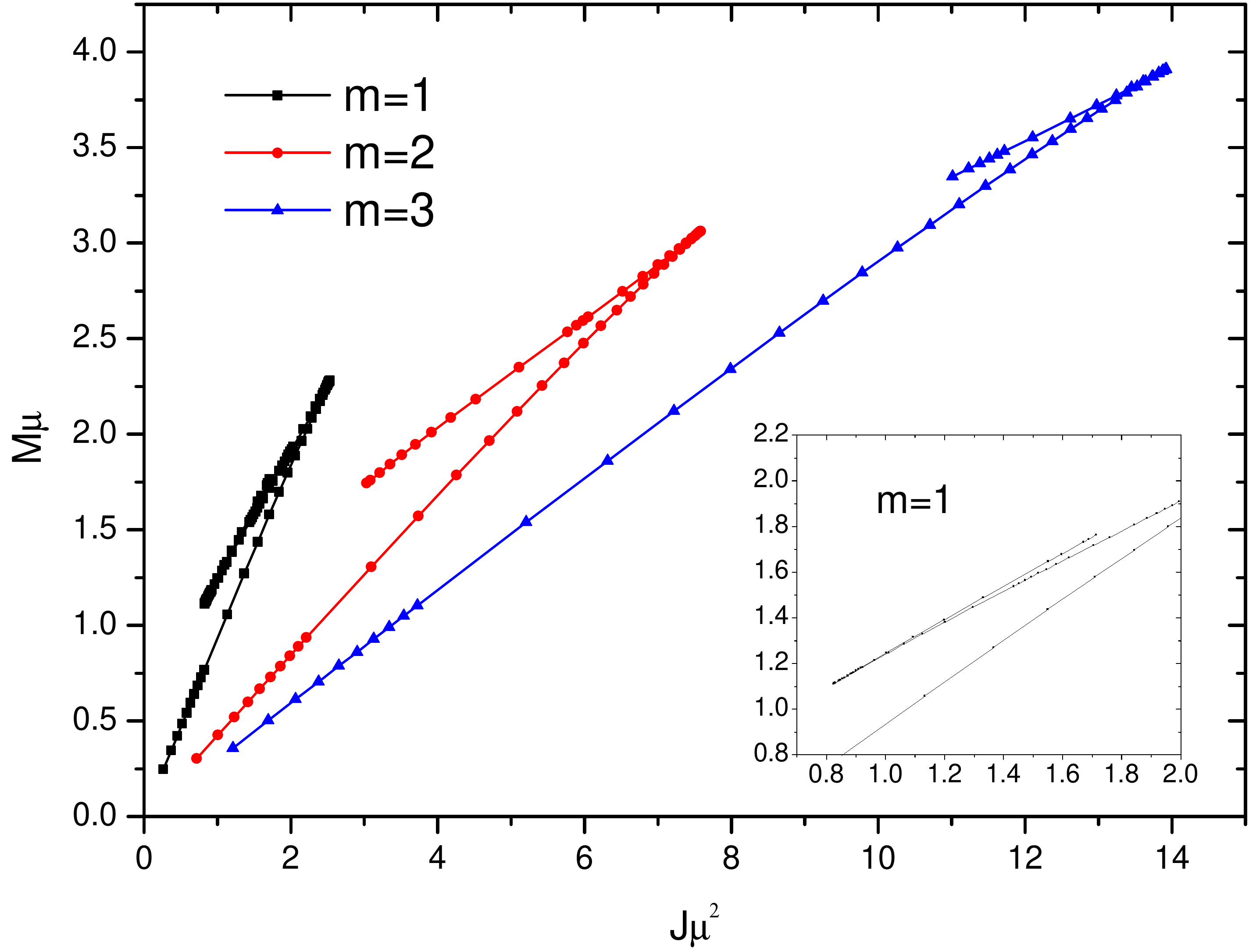}
\end{center}
\caption{The physical property  of the rotating   $n_\theta=1$  excited state of boson stars with $m=1,~2,~3$.  Left panel: the BS mass  $M$ as a function of the  frequency  $\omega$.
Right panel: the BS mass  $M$  as a function of the angular momentum $J$.}
\label{f-b2}
\end{figure}

After  obtaining  the numerical solution of rotating excited state boson stars,  in the left panel of  Fig. \ref{f-b2} we  show   the mass $M$ as a function of the  frequency  $\omega$  with the azimuthal harmonic index $m=1$ (black  line), 2  (red  line), 3 (blue  line), respectively,  and  plot  the mass $M$  as a function of the angular momentum   $J$   for the corresponding  values of  $m$ in the right panel of  Fig. \ref{f-b2}.
From the graphics, it is obvious that  there exists excited state  boson star   for $\omega < \mu$ and  the excited state solution is still  bound state.

  In $M-\omega$  diagram, the fitting curve   emanating  from the point ($\omega=\mu$, $M=0$)
moves away as it revolves around a central region of the diagram.
In contrast to the ground boson stars in Ref.  \cite{Herdeiro:2014goa} and the excited boson stars with $n_r=1$, the  excited sets of solutions with $n_\theta=1$
do not exhibit a spiraling behavior.   For example,
for $m=1$, we can see that  with the decrease of the  frequency  $\omega$, the mass of boson stars begins to increase
and then decrease until a minimum value  of $\omega$. Further,  below  the minimum value we obtain a second branch of excited boson stars with lower mass.   So far these data show a similar behavior  as the  ground state. However,  above the  minimum value  of $\omega$ we
obtain a third set of unstable solutions with larger  mass,  where  the curve
 does not spiral into the center.  It is noted that
   in contrast to the behavior of  the curve   in the left panel of  Fig. \ref{f-12}, we see that  the mass of the  excited  solutions with $n_\theta=1$   is more heavier than  the case of  $n_r=1$,    while  the minimum value  of $\omega$   of the  excited state with $n_\theta=1$  is  smaller than the case of  $n_r=1$.

 We exhibit in the  right panel of Fig.  \ref{f-b2}
how  the BS mass  $M$ varies as a function of  the angular momentum $J$  with the   azimuthal harmonic index  $m = 1$ (black line), $m = 2$ (red line), and $m = 3$ (blue line), respectively.  These three  zigzag  patterns are similar to the case of  $n_\theta=1$ in the right panel of Fig. \ref{f-12},  and in  the inset we show the detail of the zigzag curve with $m=1$.

\subsubsection{Black hole}
Let us now study the
physical properties of rotating   excited hairy black hole
solutions with $n_\theta =1 $.
In the  left panel of Fig. \ref{f-1s}, we  show   the BH mass $M$ as a function
of  the  frequency  $\omega$  with     several radii of event horizon      in  $m=1$ mode.    In the graph, the curve of the black colour represents  the excited boson star with $n_\theta =1 $,
and the solution  domain of Kerr BHs is indicated by the orange shaded area. It  is interesting to note that these families of  excited hairy black hole
solutions with $n_\theta =1 $  do not present  the spiraling behavior  as that of  $n_r =1 $  in the left panel of Fig. \ref{bh-2}.
According to  the dependence of mass on the angular momentum, the curves  can be divided into two kinds as follows:

\begin{figure}[h!]
\begin{center}
\includegraphics[height=.25\textheight,width=.34\textheight, angle =0]{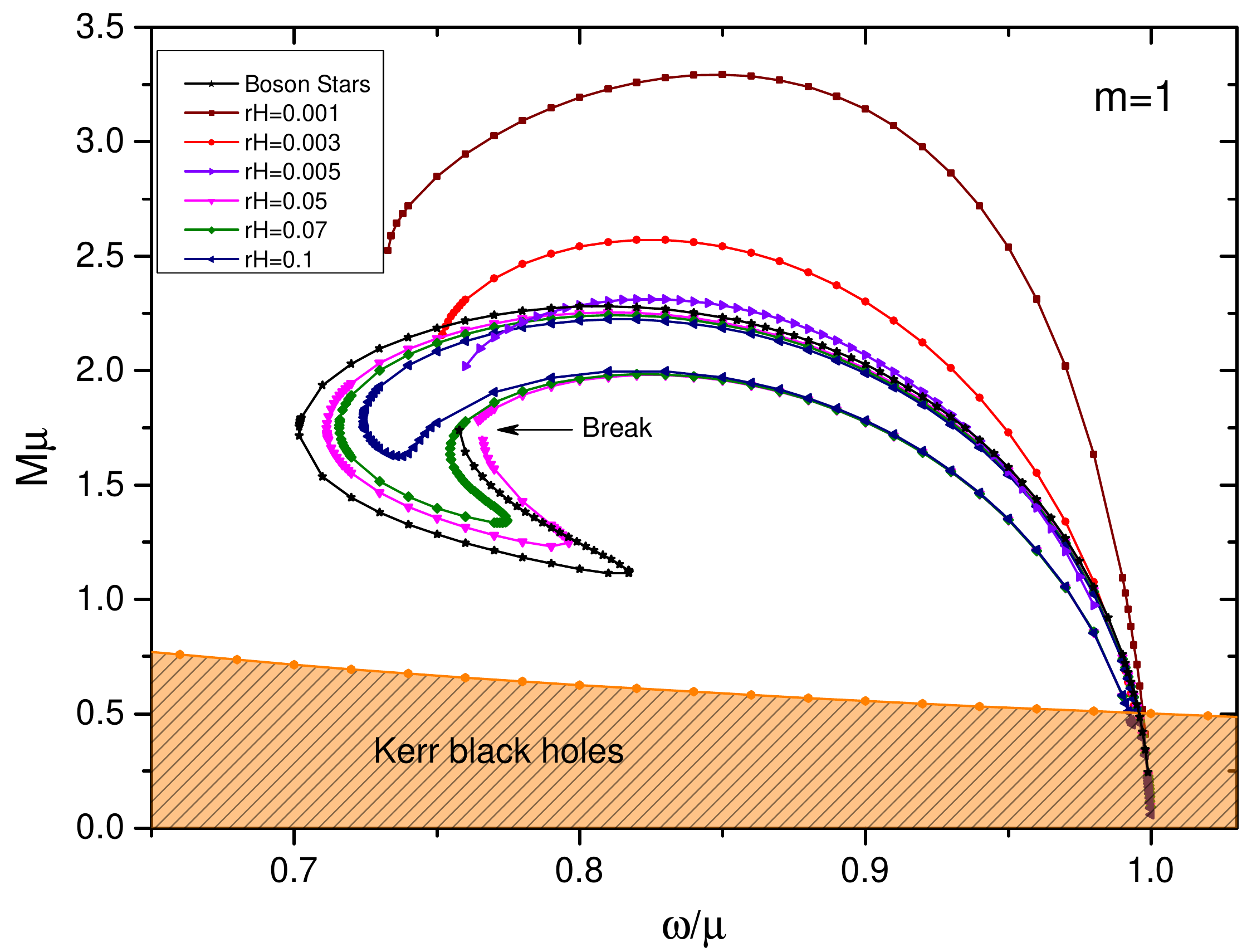}
\includegraphics[height=.25\textheight,width=.34\textheight, angle =0]{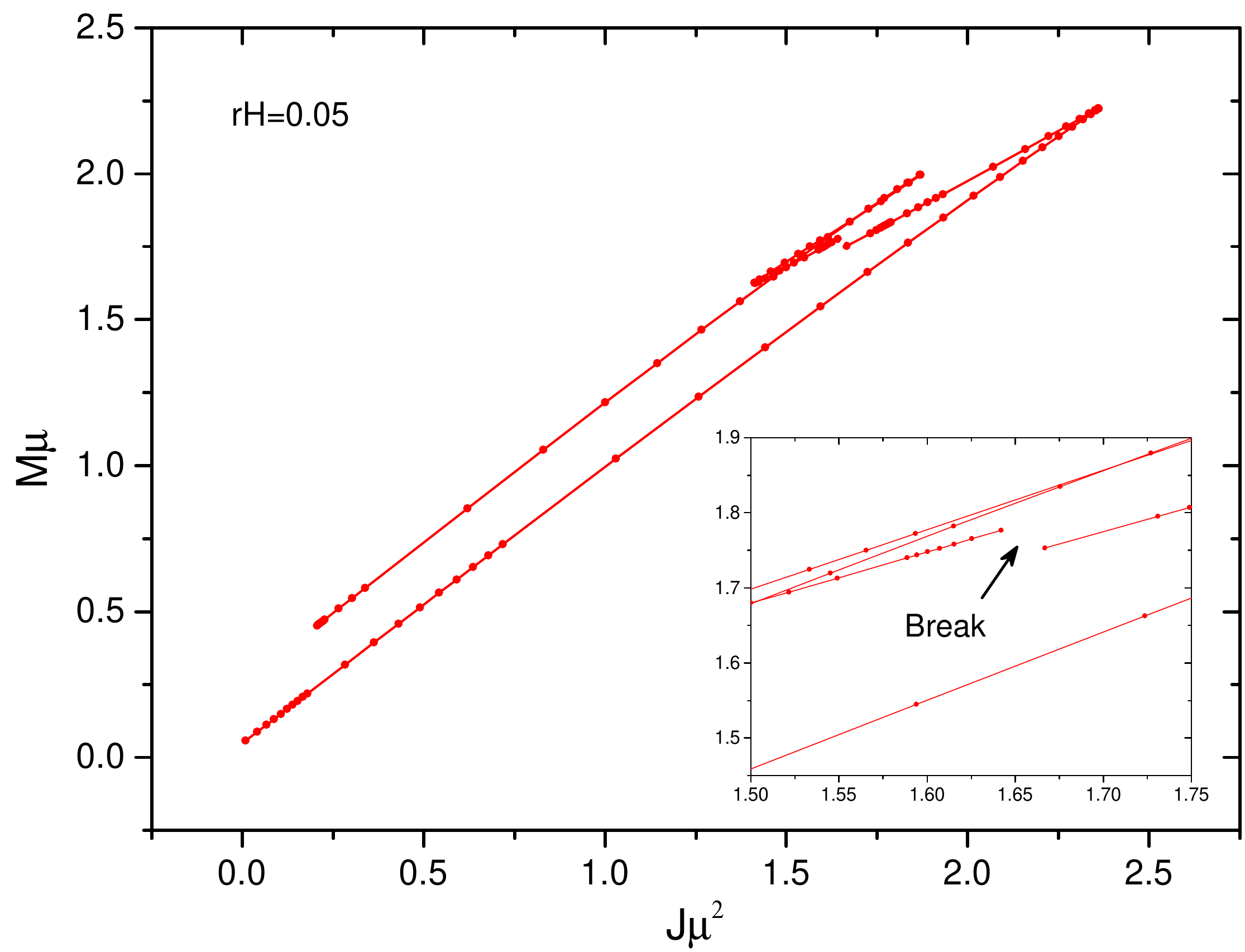}
\end{center}
\caption{The physical property  of  the   $n_\theta=1$  excited state of Kerr black holes with $m=1$ scalar hair.  Left graph: the BH mass  $M$ as a function of the  frequency  $\omega$  with  the event horizon $r_H=0.001,  0.003,  0.005,  0.05,  0.07$  and  $0.1$,   respectively.
Right graph: the BH mass  $M$  as a function of the angular momentum $J$  with  the event horizon $r_H=0.05$.}
\label{f-1s}
\end{figure}

\begin{enumerate}
\item \emph{ Closed loop:}
For   $r_H=0.1$ and 0.07, the curves  of the radially excited KBHsSH originally start from the maximal
frequency $\omega=\mu$ at the Kerr BH vacuum, and then reach a minimal
 frequency. Further increasing $\omega$,  the mass begins to decrease until a  new maximum value  of $\omega$   and then
 the curves   turn round to another minimal
 frequency,
    and finally end  at the maximal
frequency at the vacuum solution.
So,  the set of   curves
form a closed loop which is similar to the $n_\theta=1$  excited KBHsSH curve  in the left panel of
Fig. \ref{bh-2}.
\item\emph{ Open  loop: }
For the small  value of the event horizon  $r_H=0.05$ ( the pink line),  the curve is not
continuous and has a  break point.  The similar behaviour also occurs for the curves with   $r_H= 0.005,  0.003$,  and $0.001$.
Note that,  in order to avoid overlapping with  the black curve of  boson star,  we only show  the part of these curves.
 \end{enumerate}

The above two kinds of solutions with $n_\theta=1$ are very different from that of the ground state.
In the   left panel, we  only exhibit the solution of the lowest $m$
modes,  and the higher values of $m$  also have similar properties.
Furthermore,  in the right panel of Fig. \ref{f-1s}, we  plot  the mass $M$  as a function of the angular momentum $J$   for $r_H=0.05$.
We can  see the  zigzag  pattern also has  a break point,  which just corresponds to the position   in the $M-\omega$ curve,  and more details  are  shown in the inset.

\begin{figure}[h!]
\begin{center}
\includegraphics[height=.25\textheight,width=.34\textheight, angle =0]{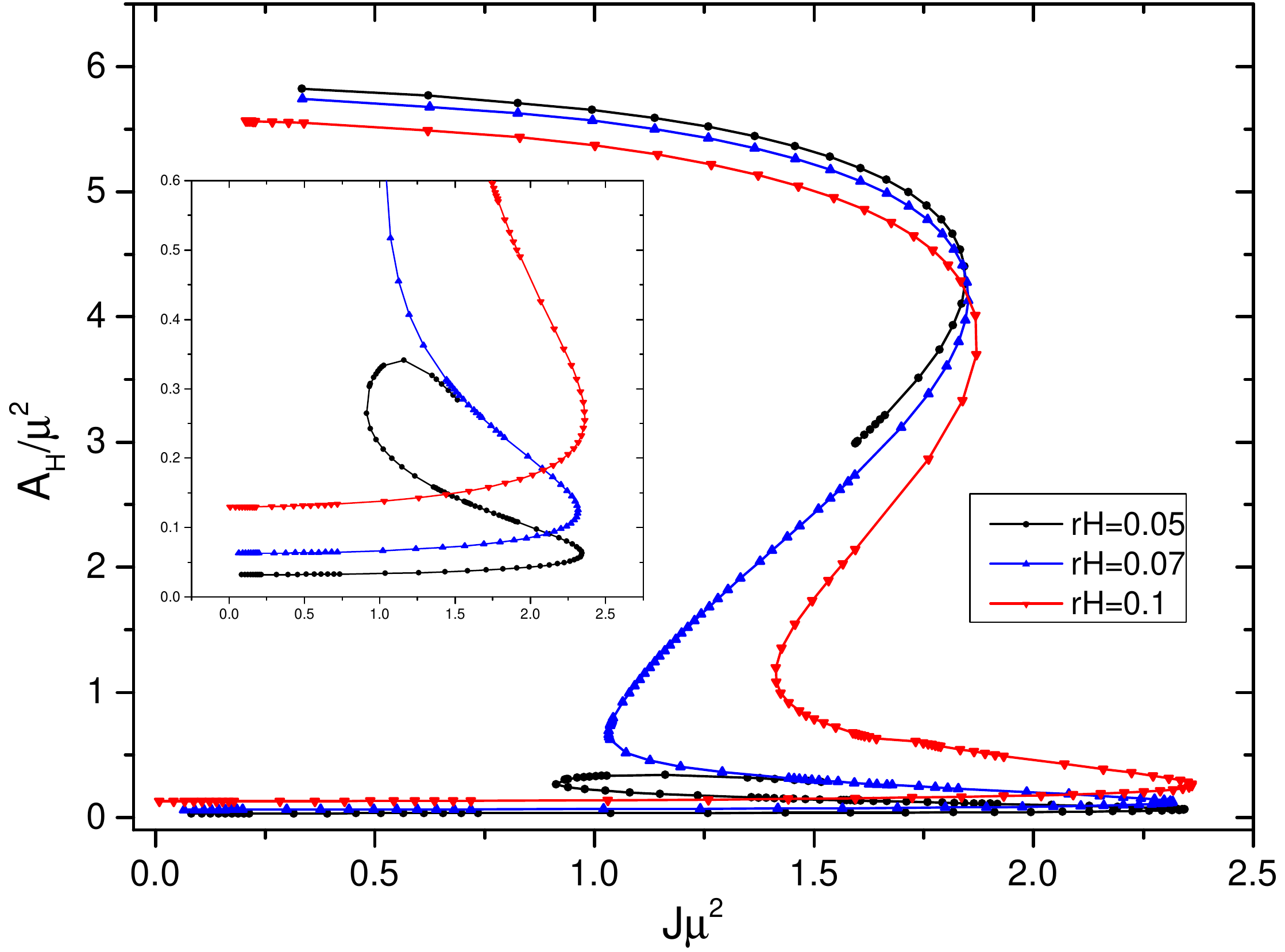}
\includegraphics[height=.26\textheight,width=.34\textheight, angle =0]{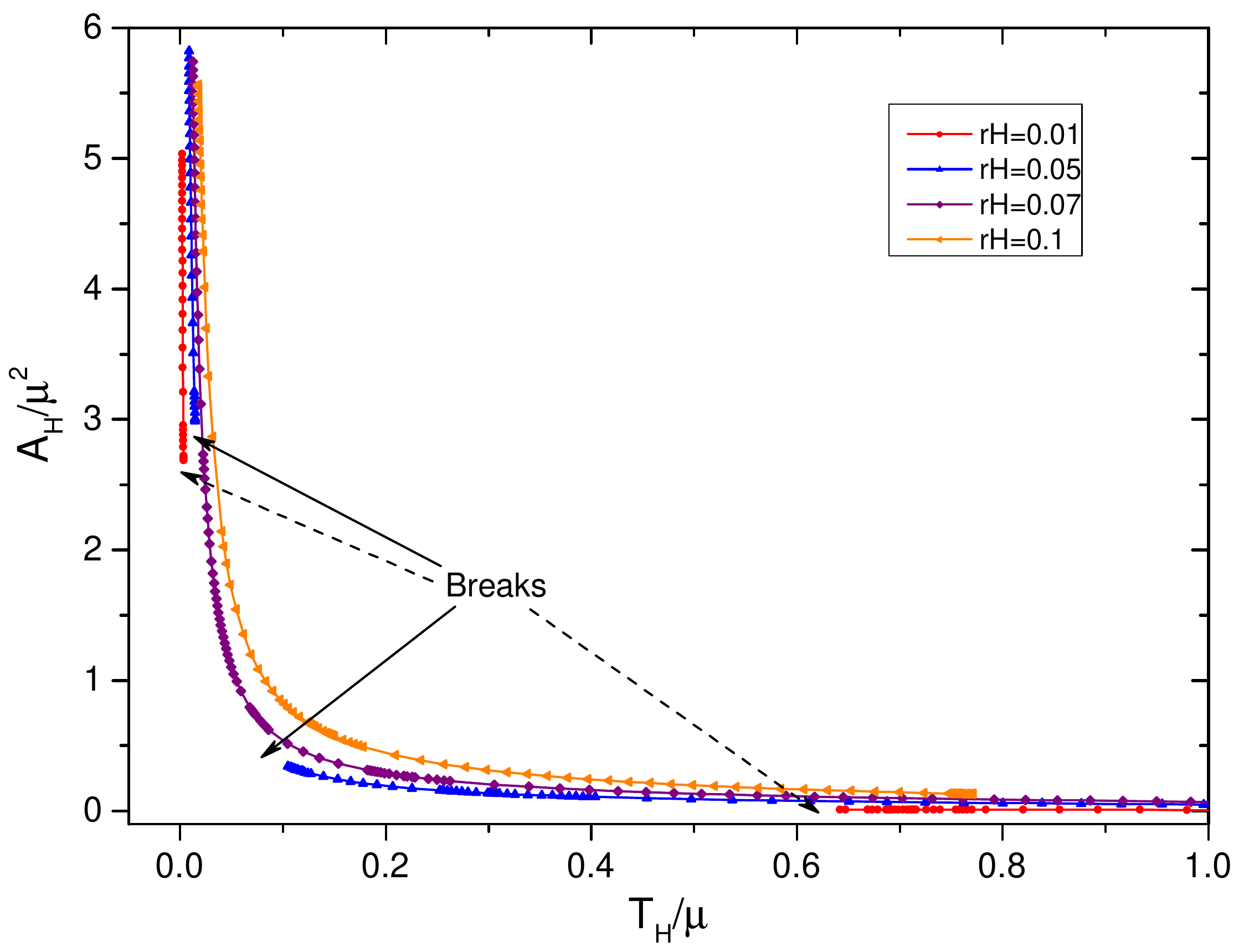}
\end{center}
\caption{The surface area  $A_H$  of  the   $n_\theta=1$  excited state of Kerr black holes with $m=1$ scalar hair.  Left panel:  the surface area   as a function of the angular momentum $J$  with  the event horizon $r_H=0.05$ (black), $ 0.07$ (blue),  and  $0.1$ (red),   respectively.
Right panel: the surface area   as a function of the  Hawking temperature $T_H$ with  the event horizon  $r_H = 0.01,   0.05,  0.07$,  and  0.1,   respectively.}
\label{f-1sss}
\end{figure}
The relation between the event horizon   $A_H$ and    angular momentum $J$  is shown in the left panel of  in Fig. \ref{f-1sss},
where we plot  several typical examples for our numerical results
by taking $r_H=0.05, 0.07$ and 0.1,  respectively.
 We   find that hairy black hole solutions  with $r_H=0.07$ and 0.1 have similar behavior as those given in the left panel of  Fig. \ref{bh-3}. However,  the black curve with  $r_H=0.05$   is not
continuous and has a  break point.
 There exist two   separate  branches   of   how  the area $A_H$ of the event horizon  varies with  $J$.
 Indeed, this behavior is in good agreement with the numerical results in Fig. \ref{f-1s}.
More details are shown as an inset plot in the left panel of
Fig. \ref{f-1sss}.   In the right panel,   the event horizon   $A_H$ as a function of   the temperature $T_H$ is shown
by taking $r_H=0.01, 0.05,0.07$ and 0.1, respectively.  The curves with $r_H=0.07$ and 0.1 are continuum, while the ones with
 $r_H=0.01$ and 0.05  have a break point. Moreover, with decreasing $r_H$, the range of break   becomes larger.

\section{Conclusion}\label{sec4}
In this paper,  we
have analyzed  the model of  (3+1)-dimensional Einstein gravity coupled to a complex,  massive  scalar field and numerically constructed the  solutions of
 rotating compact objects with excited scalar hair, including boson stars and black holes.
Comparing with the ground state solution in Ref.  \cite{Herdeiro:2014goa},  we found that the first-excited   Kerr BHs with scalar hair  have  two   types of nodes,  radial $n_r=1$  and angular  $n_\theta=1$ nodes.
In the former case the curves of the mass  versus the
 frequency form nontrivial loops, starting from and
ending at the trivial solution at   $\omega_{max}$.
 In the latter case  the curves  can be divided into two types: the closed and open loops.  For a larger  value of the event horizon radius,  the curve forms a closed loop.  While for a small  one,  the curve is not
continuous and has a  break point.   In addition, the range of break point
   decreases with  $r_H$.
It is notable that  there is  a similar situation
 in atomic theory and quantum mechanics,  for example, the first excited state of hydrogen has an electron in the 2s-orbital or 2p-orbital,
 which corresponds to the radial or angel node, respectively.
 From the numerical results, we can see that the mass of the excited KBHsSH solution is  heavier than  the ground state. Moreover, the state with
 $n_\theta=1$ has higher mass level than the one with  $n_r=1$.

 There are several  interesting extensions of our work.
First, we have studied  the first excited   Kerr BH with a free scalar hair, 
 we would like to investigate how  self-interactions of the scalar field  affects the excited
Kerr black holes with scalar hair inspired by  the work \cite{Herdeiro:2015tia}.
The second  extension of our study is to  construct generalized multi-scalar hair configurations, where two coexisting states of the scalar field are
presented, including ground and excited states. The first step in this
direction was done in \cite{Bernal:2009zy},   where the rotating boson star
with coexisting ground and excited  states   was constructed.
Finally, we are planning to study the model of
the Einstein-complex-Proca model and construct the  excited   Kerr BHs with Proca hair  in future work.

\section*{Acknowledgement}
YQW would like to thank  Hao Wei  and Li Li   for  helpful discussion.   Some
computations were performed on the   Shared Memory system at  Institute of Computational Physics and Complex Systems in Lanzhou University. This work was supported by the  Natural Science Foundation of China (Grants No. 11675064,  No. 11522541 and No. 11875175), and the Fundamental Research Funds for the Central Universities (Grants No. lzujbky-2017-182, No. lzujbky2017-it69 and  No. lzujbky-2018-k11).

\providecommand{\href}[2]{#2}\begingroup\raggedright
\endgroup

\end{document}